\documentclass{article}
\usepackage{spconf,amsmath,graphicx}
\usepackage{amssymb}
\usepackage[hidelinks]{hyperref}
\usepackage[backend=biber,
            style=ieee,
            doi=false,
            isbn=false,
            url=false]{biblatex}
\usepackage{tikz}
\usetikzlibrary{dsp,positioning}

\newcommand{\dif}{\, \mathrm{d}}

\newcommand{\bftab}{\fontseries{b}\selectfont}
\usepackage[keeplastbox]{flushend}

\usepackage{graphics}

\usepackage{acronym}
\newacro{SNR}{signal-to-noise ratio}
\newacro{STFT}{short-time Fourier transform}
\newacro{STSA}{short-time spectral amplitude}
\newacro{MMSE}{minimum mean squared error}
\newacro{MSE}{mean squared error}
\newacro{TCN}{temporal convolution network}
\newacro{MAP}{\emph{maximum a posteriori}}
\newacro{A-MAP}{approximate MAP}
\newacro{ML}{maximum likelihood}
\newacro{iSTFT}{inverse STFT}
\newacro{SI-SDR}{scale-invariant signal-to-distortion ratio}
\newacro{DNS}{Deep Noise Suppression}
\newacro{ESTOI}{extended short-time objective intelligibility}
\newacro{PESQ}{perceptual evaluation of speech quality}
\newacro{DNN}{deep neural network}
\newacro{POLQA}{perceptual objective listening quality  analysis}

\DeclareMathOperator*{\argmin}{argmin} %

\usepackage[all=normal,floats=tight,charwidths=tight, mathspacing =tight, paragraphs=tight]{savetrees}
\title{Integrating statistical uncertainty into neural network-based speech enhancement}

\name{Huajian Fang$^{1,2}$, Tal Peer$^1$,  Stefan Wermter$^{2}$, Timo Gerkmann$^1$\thanks{This work has been funded by the Deutsche Forschungsgemeinschaft (DFG, German Research Foundation) project number 247465126}}
\address{
  $^1$Signal Processing (SP), Universität Hamburg, Germany\\
  $^2$Knowledge Technology (WTM), Universität Hamburg, Germany\\
  \texttt{\{huajian.fang,tal.peer,stefan.wermter,timo.gerkmann\}@uni-hamburg.de}}

\begin{document}
\ninept
\maketitle
\begin{abstract}
Speech enhancement in the time-frequency domain is often performed by estimating a multiplicative mask to extract clean speech. However, most neural network-based methods perform point estimation, i.e., their output consists of a single mask. In this paper, we study the benefits of modeling uncertainty in neural network-based speech enhancement. For this, our neural network is trained to map a noisy spectrogram to the Wiener filter and its associated variance, which quantifies uncertainty, based on the \emph{maximum a posteriori}~(MAP) inference of spectral coefficients. By estimating the distribution instead of the point estimate, one can model the uncertainty associated with each estimate. We further propose to use the estimated Wiener filter and its uncertainty to build an approximate MAP~(A-MAP) estimator of spectral magnitudes, which in turn is combined with the MAP inference of spectral coefficients to form a hybrid loss function to jointly reinforce the estimation. Experimental results on different datasets show that the proposed method can not only capture the uncertainty associated with the estimated filters, but also yield a higher enhancement performance over comparable models that do not take uncertainty into account. 
\end{abstract}
\section{Introduction}
 \vspace{-0.03cm}
\label{sec:intro}
Single-channel speech enhancement algorithms typically operate in the \ac{STFT} domain~\cite{timowienerfiltering2018, noisereducitonsurvey, timo2012}. The Gaussian statistical model in the \ac{STFT} domain has been shown to be effective~\cite{timowienerfiltering2018, ephraim1984speech}. Given the assumption that the complex-valued speech and noise coefficients are uncorrelated and Gaussian-distributed with zero mean, various estimators have been derived, such as the Wiener filter and the \ac{STSA} estimator~\cite{timowienerfiltering2018,ephraim1984speech, wolfe2003efficient}. The Wiener filter, which is optimal in the \ac{MMSE} sense, requires estimation of speech and noise variances. This can be achieved by various signal processing estimators with varying degrees of success for different signal characteristics~\cite{timo2011waspaa, noisereducitonsurvey, timowienerfiltering2018, carbajal2021guided, huajian2021noiseaware, richter2020speech, simonvae, bandovae}.

Recently, \acp{DNN} have been successfully applied to speech enhancement and regularly show an improved performance over classical methods~\cite{bandovae, simonvae, wang2018supervised, rehr2021snr}. Among the DNN-based approaches relevant to this work are deep generative models (e.g., variational autoencoder) and supervised masking approaches. Generative models estimate the clean speech distribution and subsequently combine it with a separate noise model to construct a point estimate of a noise-removing mask (Wiener filter)~\cite{bandovae, simonvae}. In contrast, typical supervised learning approaches are trained on pairs of noisy and clean speech samples and directly estimate a time-frequency mask that aims at reducing noise interference with minimal speech distortion given a noisy mixture, using a suitable loss function (e.g., \ac{MSE})~\cite{wang2018supervised, rehr2021snr}. However, the supervised approaches often learn the mapping between noisy and clean speech blindly and output a single point estimate without guarantee or measure of its accuracy. In this work we focus on adding an uncertainty measure to a supervised method by estimating the speech posterior distribution, instead of only its mean. Note that while this is conceptually related to the generative approach, in this case we do not estimate the clean speech prior distribution, but rather the posterior distribution of clean speech given a noisy mixture.

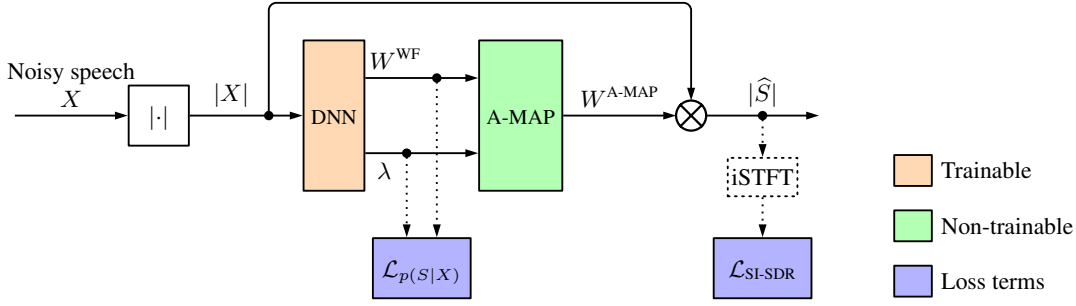
\begin{figure*}[!ht]
  \centering
      \tikzstyle{normalnode}=[dspsquare, inner sep=1mm]%
    \tikzstyle{bignode}=[normalnode, minimum height=2cm, align=center]
    \tikzstyle{smallnode}=[normalnode]
    \tikzstyle{legendnode}=[normalnode, minimum height=0.2cm, minimum width=0.5cm, inner sep=0]
    \tikzstyle{loss}=[normalnode, fill=blue!30, minimum width=1.3cm]
    \tikzstyle{connarrow}=[dspconn]
    \begin{tikzpicture}[node distance=1cm and 1.5cm]
        \node[smallnode] (Xmag) {$\left| \cdot \right|$};
        \node[left=of Xmag] (Xin) {};
        \node[bignode, fill=orange!30, right=of Xmag] (NN) {\footnotesize DNN};
        \node[bignode, fill=green!30, right=of NN] (AMAP) {\footnotesize A-MAP};
        \node[dspmixer, right=of AMAP] (mul) {};
        \node[right=of mul] (Smag_hat) {};

        \node[dspnodefull, right=1cm of Xmag] (Xmag_split) {};

        \draw[connarrow] (Xin) -- node[above, align=center] {Noisy speech\\$X$} (Xmag);
        \draw[connarrow] (Xmag) -- node[above, xshift=-0.2cm] {$\left| X \right|$} (NN);
        \draw[connarrow] ([yshift=0.5cm]NN.east) -- node[above,xshift=-0.35cm] {$W^{\text{WF}}$} node[dspnodefull,xshift=0.2cm] (w_split) {} ([yshift=0.5cm]AMAP.west);
        \draw[connarrow] ([yshift=-0.5cm]NN.east) -- node[below,xshift=-0.5cm] {$\lambda$} node[dspnodefull,xshift=-0.2cm] (lambda_split) {} node[] (lambda_dummy) {} ([yshift=-0.5cm]AMAP.west);
        \draw[connarrow] (AMAP) -- node[above] {$W^{\text{A-MAP}}$} (mul);
        \draw[connarrow] (mul) -- node[above] {$| \widehat{S} |$} node[dspnodefull] (Smag_hat_split) {} (Smag_hat);
        
        \draw[connarrow, rounded corners=1mm] (Xmag_split) -- +(0, 1.5) -| (mul);

        \node[loss, below=of lambda_dummy] (unc_loss) {$\mathcal{L}_{p(S|X
        )}$};
        \draw[connarrow, dotted] (w_split) -- ([xshift=0.2cm]unc_loss.north);
        \draw[connarrow, dotted] (lambda_split) -- ([xshift=-0.2cm]unc_loss.north);

        \node[loss] at (Smag_hat_split |- unc_loss) (sisdr_loss) {$\mathcal{L}_{\text{SI-SDR}}$};
        \node[legendnode, dotted, below=0.5cm of Smag_hat_split, inner sep=0.5mm] (istft) {iSTFT};
        \draw[connarrow, dotted] (Smag_hat_split) -- (istft.north);
        \draw[connarrow, dotted] (istft.south) -- (sisdr_loss.north);

        \node[legendnode, fill=orange!30, label=right:{\small Trainable}, xshift=2cm, yshift=1.3cm] at (sisdr_loss |- unc_loss) (legend_train) {};
        \node[legendnode, fill=green!30, label=right:{\small Non-trainable}, below=0.3cm of legend_train] (legend_fixed) {};.
        \node[legendnode, fill=blue!30, label=right:{\small Loss terms}, below=0.3cm of legend_fixed] (legend_loss) {};

    \end{tikzpicture}
\caption{Block diagram of the neural network-based uncertainty estimation. The neural network is trained  according to the proposed hybrid loss function. }
\label{fig:uncertainty_diagram}
\end{figure*} 
Uncertainty modeling based on neural networks has been actively studied in e.g., computer vision~\cite{uncertaintyincvalex2017}. Inspired by this, here we propose a hybrid loss function to capture uncertainty associated with the estimated Wiener filter in the neural network-based speech enhancement algorithm, as depicted in Fig.~\ref{fig:uncertainty_diagram}. More specifically, we propose to train a neural network to predict the Wiener filter and its variance, which quantifies the uncertainty, based on the \ac{MAP} inference of complex spectral coefficients, such that full Gaussian posterior distribution can be estimated. To regularize the variance estimation, we build an \ac{A-MAP} estimator of spectral magnitudes using the estimated Wiener filter and uncertainty, which is in turn used together with the \ac{MAP} inference of spectral coefficients to form a hybrid loss function. Experimental results show the effectiveness of the proposed approach in capturing uncertainty. Furthermore, the \ac{A-MAP} estimator based on the estimated Wiener filter and its associated uncertainty results in improved speech enhancement performance.

 \vspace{-0.075cm}
\section{Signal model}
\label{sec:model}
 \vspace{-0.03cm}
We consider the speech enhancement problem in the single microphone case with additive noise. The noisy signal $x$ can be transformed into the time-frequency domain using the \ac{STFT}: 
\begin{equation}
X_{ft} = S_{ft} + N_{ft} \, ,
\label{eqn:timemodel}
\end{equation}
where $X_{ft}$, $S_{ft}$, and $N_{ft}$ are complex noisy speech coefficients, complex clean speech coefficients, and complex noise coefficients, respectively. The frequency and frame indices are given by $f \in \{1,2,\cdots, F\}$ and $t \in \{1,2,\cdots, T\}$, where $F$ denotes the number of frequency bins, and $T$ represents the number of time frames. Furthermore, we assume a Gaussian statistical model, where the speech and noise coefficients are uncorrelated and follow a circularly symmetric complex Gaussian distribution with zero mean, i.e., 
 \begin{equation}
 \label{eq:prior}
S_{ft} \sim \mathcal{N}_\mathbb{C}(0,\,\sigma^{2}_{s,ft}), \hspace{0.3cm}
N_{ft} \sim \mathcal{N}_\mathbb{C}(0,\,\sigma^{2}_{n,ft}) \, ,
 \end{equation}
where $\sigma^{2}_{s,ft}$ and  $\sigma^{2}_{n,ft}$ represent the variances of speech and noise, respectively. The likelihood $p(X_{ft}|S_{ft})$ follows a complex Gaussian distribution with mean $S_{ft}$ and variance $\sigma_{n, ft}^2$, given by
\begin{equation}
    \label{eq:likelihood}
    p(X_{ft}|S_{ft}) = \frac{1}{\pi \sigma^2_{n,ft}} \exp\left(-\frac{|X_{ft}-S_{ft}|^2}{\sigma^2_{n,ft}}\right) \, .
\end{equation}
Given the speech prior in~\eqref{eq:prior} and the likelihood in~\eqref{eq:likelihood}, we can apply Bayes' theorem to find the speech posterior distribution, which is complex Gaussian of the form
\begin{equation}
    \label{eqn:posteriorcoplex}
    p(S_{ft}|X_{ft}) = \frac{1}{\pi \lambda_{ft}} \exp{\left(-\frac{|S_{ft} - W^{\text{WF}}_{ft}X_{ft}|^2}{\lambda_{ft}}\right)} \, ,
\end{equation}
where $W^{\text{WF}}_{ft} = \frac{\sigma_{s,ft}^2}{\sigma_{s,ft}^2 + \sigma_{n,ft}^2}$ is the \emph{Wiener filter} and $\lambda_{ft} = \frac{\sigma_{s,ft}^2\sigma_{n,ft}^2}{\sigma_{s,ft}^2 + \sigma_{n,ft}^2}$ is the posterior's variance~\cite{timowienerfiltering2018}. The \ac{MMSE} and \ac{MAP} estimators of $S_{ft}$ under this model are both given by the \emph{Wiener filter} \cite{timowienerfiltering2018}: $\widetilde{S}_{ft} = W^{\text{WF}}_{ft} X_{ft}$. It is known that the expectation of \ac{MMSE} estimation error is closely related to the posterior variance~\cite{astudillo2009accounting}, and under the assumption of complex Gaussian distribution it corresponds directly to the variance, i.e.,
\begin{equation}
\begin{split}
    E\{|S_{ft}-\widetilde{S}_{ft}|^2\} & = \iint |S_{ft}-\widetilde{S}_{ft}|^2 \, p(S_{ft}|X_{ft})p(X_{ft})\dif S_{ft}\dif X_{ft} \\
    & = \int \lambda_{ft} \, p(X_{ft}) \dif X_{ft} = \lambda_{ft}.
\end{split}
     \label{mmse_error}
\end{equation}
The variance $\lambda_{ft}$ can be interpreted as a measure of uncertainty associated with the \ac{MMSE} estimator~\cite{timowienerfiltering2018}. In the following sections $\lambda_{ft}$ will be referred to as the (estimation) uncertainty.

 \vspace{-0.1cm}
\section{Deep Uncertainty Estimation}
\label{sec:deepuncertainty}
 \vspace{-0.05cm}
The Wiener filter can be computed for a given noisy signal by estimation of $\sigma^{2}_{s,ft}$ and $\sigma^{2}_{n,ft}$ using traditional signal processing techniques. It is, however, also possible to directly estimate $W^{\text{WF}}_{ft}$ using a \ac{DNN}.
Furthermore, if optimization is based on the posterior~\eqref{eqn:posteriorcoplex}, besides $W^{\text{WF}}_{ft}$ also the uncertainty $\lambda_{ft}$ can be estimated as previously proposed in the computer vision domain~\cite{uncertaintyincvalex2017}. Taking the negative logarithm (which does not affect the optimization problem due to monotonicity) and averaging over the time-frequency plane results in the following minimization problem:
\begin{equation}
    \begin{split}
    &\widetilde{W}^{\text{WF}}_{ft}, \widetilde{\lambda}_{ft} = \\
    &\argmin_{W^{\text{WF}}_{ft},\lambda_{ft}} \underbrace{\frac{1}{FT}\sum_{f,t} \log(\lambda_{ft})  + \frac{|S_{ft} - W^{\text{WF}}_{ft} X_{ft}|^2}{\lambda_{ft}}}_{\mathcal{L}_{p(S|X)}} \, ,   
    \end{split}
    \label{eqn:logposterior}
\end{equation}
where $\widetilde{W}_{ft}$, $\widetilde{\lambda}_{ft}$ denote estimates of the Wiener filter and its uncertainty. If we assume a constant uncertainty for all time-frequency bins, i.e., $\lambda_{ft} = \lambda^\ast$, and refrain from explicitly optimizing for $\lambda^\ast$, $\mathcal{L}_{p(S|X)}$ degenerates into the well known \ac{MSE} loss
\begin{equation}
\mathcal{L}_{\text{MSE}} = \frac{1}{FT}\sum_{f,t}|S_{ft}-W^{\text{WF}}_{ft}X_{ft}|^2 \, ,
\label{eqn:mse}
\end{equation}
which is widely used in DNN-based regression tasks, including speech enhancement~\cite{wang2018supervised, braun2021loss}. In this work we depart from the assumption of constant uncertainty. Instead, we propose to include uncertainty estimation as an additional task by training a DNN with the full negative log-posterior $\mathcal{L}_{p(S|X)}$.

It has been previously shown that modeling uncertainty by minimizing $\mathcal{L}_{p(S|X)}$ results in improvement over baselines that do not take uncertainty into account in computer vision tasks~\cite{uncertaintyincvalex2017}. However, in preliminary experiments we have observed that directly using~\eqref{eqn:logposterior} as loss function results in reduced estimation performance for the Wiener filter and is prone to overfitting. To overcome this problem, we propose an additional regularization of the loss function by incorporating the estimated uncertainty into clean speech estimation as described next.

 \vspace{-0.15cm}
\section{Joint~enhancement and uncertainty estimation}
\label{sec:proposedscheme}
Besides estimation of the Wiener filter and its uncertainty, we propose to also incorporate a subsequent speech enhancement task that explicitly uses both into the training procedure. The speech enhancement task provides additional coupling between the DNN outputs (Wiener filter and uncertainty). In this manner, the DNN is guided towards estimation of uncertainty values that are relevant to the speech enhancement task, as well as enhanced estimation of the Wiener filter.

If we consider complex coefficients with symmetric posterior~\eqref{eqn:posteriorcoplex}, the MAP and MMSE estimators both result directly in the Wiener filter $W^{\text{WF}}_{ft}$ and do not require an uncertainty estimate. However, this changes if we consider spectral magnitude estimation. The magnitude posterior $p(|S_{ft}|\:|X_{ft})$, found by integrating the phase out of~\eqref{eqn:posteriorcoplex}, follows a Rician distribution~\cite{wolfe2003efficient}
\begin{equation}
\begin{split}
    &p(|S_{ft}|\:|X_{ft}) =\\ &\frac{2|S_{ft}|}{\lambda_{ft}} \exp\left(-\frac{|S_{ft}|^2+(W_{ft}^{\text{WF}})^2|X_{ft}|^2}{\lambda_{ft}}\right)\mathit{I_0}\left(\frac{2|X_{ft}|\,|S_{ft}|W^{\text{WF}}_{ft}}{\lambda_{ft}}\right)\,,
\end{split}
\end{equation}
where $\mathit{I_0}(\cdot)$ is the modified zeroth-order Bessel function of the first kind.

In order to compute the MAP estimate for the spectral magnitude, one needs to find the mode of the Rician distribution, which is difficult to do analytically. However, one may approximate it with a simple closed-form expression~\cite{wolfe2003efficient}:
\begin{equation}
    \label{eqn:approximated_map}
    \begin{split}
    |\widehat{S}_{ft}| &\approx W^{\text{A-MAP}}_{ft}|X_{ft}|\\
    &= \left(\frac{1}{2}W^{\text{WF}}_{ft} + \sqrt{\left(\frac{1}{2}W^{\text{WF}}_{ft}\right)^2 + \frac{\lambda_{ft}}{4|X_{ft}|^2}}\right) |X_{ft}| \, ,        
    \end{split}
\end{equation}
where $|\widehat{S}_{ft}|$ is an estimate of the clean spectral magnitude $|S_{ft}|$ using the \ac{A-MAP} estimator of spectral magnitudes $W^{\text{A-MAP}}_{ft}$. It can be seen that the estimator $W^{\text{A-MAP}}_{ft}$ makes use of both the Wiener filter $W^{\text{WF}}_{ft}$ and the associated uncertainty $\lambda_{ft}$. An estimate of the time-domain clean speech signal, denoted as $\widehat{s}$, is then obtained by combining the estimated magnitude $|\widehat{S}_{ft}|$ with the noisy phase, followed by the \ac{iSTFT}. The estimated time-domain signal is then used to compute the negative \ac{SI-SDR} metric \cite{le2019sdr}: 
\begin{equation}
\label{eq:sisdr}
    \mathcal{L}_{\text{SI-SDR}} = -10\log_{10}\left(\frac{||\alpha s||^2}{||\alpha s - \widehat{s}||^2}\right)\, , \quad \alpha = \frac{\widehat{s}^{T}s}{||s||^2}\, ,
\end{equation}
which is in turn used as an additional term in the loss function that forces the speech estimate (computed with $W^{\text{A-MAP}}_{ft}$) to be similar to the clean target $s$.

Finally, we propose to combine the \ac{SI-SDR} loss $\mathcal{L}_{\text{SI-SDR}}$ with the negative log-posterior $\mathcal{L}_{p(S|X)}$ given in~\eqref{eqn:logposterior}, and train the neural network using a hybrid loss  
\begin{equation}
    \mathcal{L} = \beta \mathcal{L}_{p(S|X)} + (1-\beta) \mathcal{L}_{\text{SI-SDR}}\, ,
    \label{proposedloss}
\end{equation}
with the weighting factor $\beta \in [0,1]$ as the hyperparameter. By explicitly using the estimated uncertainty for the speech enhancement task, the hybrid loss guides both mean and variance estimation to improve speech enhancement performance. An overview of this approach is  depicted in Fig.~\ref{fig:uncertainty_diagram}.
\vspace{-0.15cm}
\section{Experimental setting}
\label{sec:experimetnal setting}

 \vspace{-0.15cm}
\subsection{Dataset}
For training we use the \ac{DNS} Challenge dataset~\cite{reddy2020interspeech}, which includes a large amount of synthesized noisy and clean speech pairs. We randomly sample a subset of 100 hours with \acp{SNR} uniformly distributed between -5~dB and 20~dB. The data are randomly split into training and validation sets (80\% and 20\% respectively).

Evaluation was performed on the synthetic test set without reverberation from \ac{DNS} Challenge. Noisy signals are generated by mixing clean speech signals from~\cite{pirker2011pitch} with noise clips sampled from 12 noise categories~\cite{reddy2020interspeech}, with \acp{SNR} uniformly drawn from 0~dB to 25~dB. To examine performance across different datasets, we additionally synthesized another test dataset using clean speech signals from the \texttt{si\_et\_05} subset of the WSJ0~\cite{garofolo1993csr} dataset and four types of noise signals from CHiME~\cite{chime3dataset} (\texttt{cafe}, \texttt{street}, \texttt{pedestrian}, and \texttt{bus}) with \acp{SNR} randomly sampled from \{-10~dB, -5~dB, 0~dB, 5~dB, 10~dB\}. A few samples are dropped due to the clipping effect in the mixing processing, and finally, this results in a test dataset of 623 files.

\begin{figure}[ht!]
\begin{minipage}[b]{0.5\linewidth}
  \centering
  \centerline{\includegraphics[width=4cm, height=3.5cm]{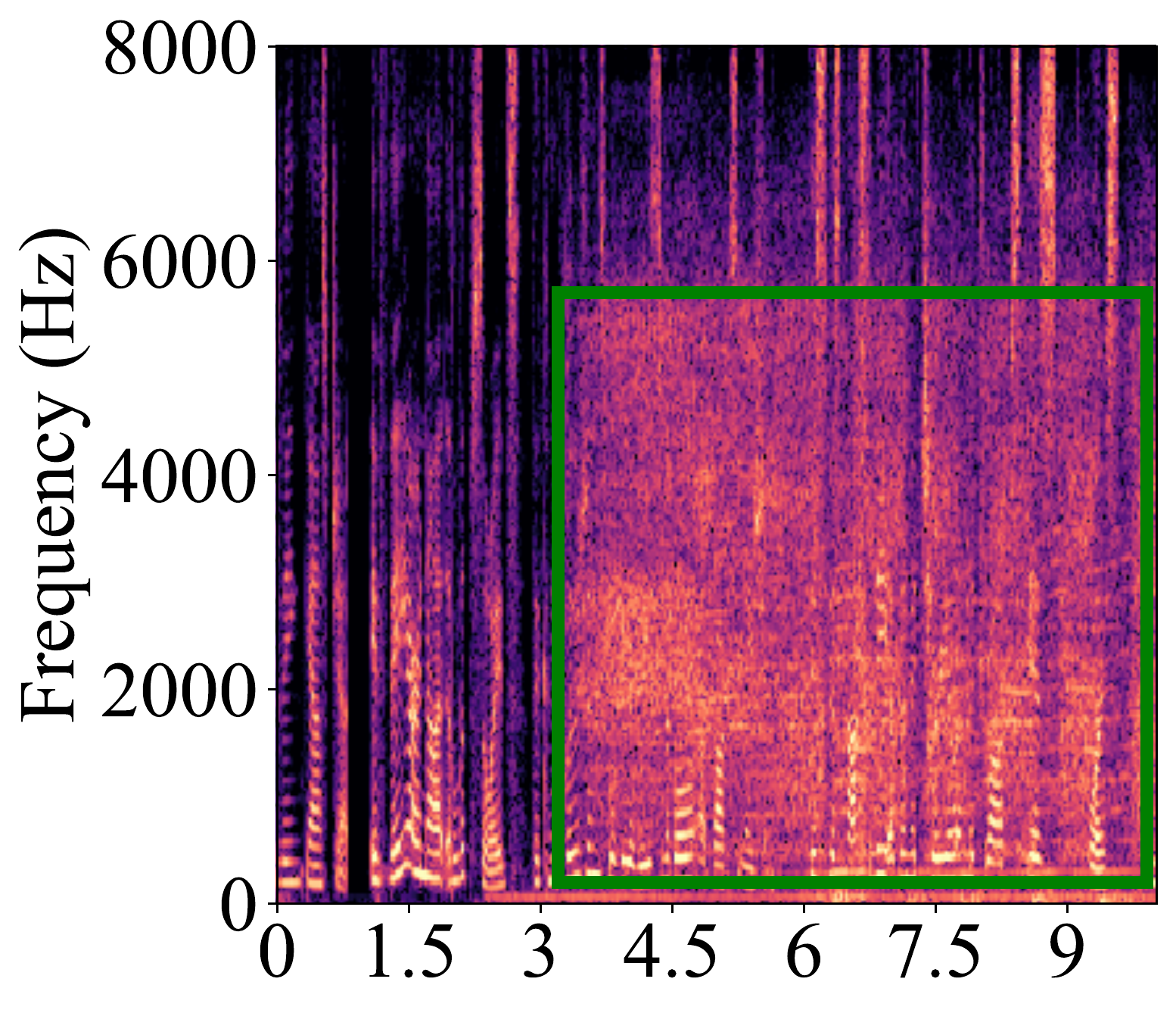}}
  \centerline{(a) Noisy}
\end{minipage}
\begin{minipage}[b]{0.51\linewidth}
  \centering
  \centerline{\includegraphics[width=4.7
  cm, height=3.5cm]{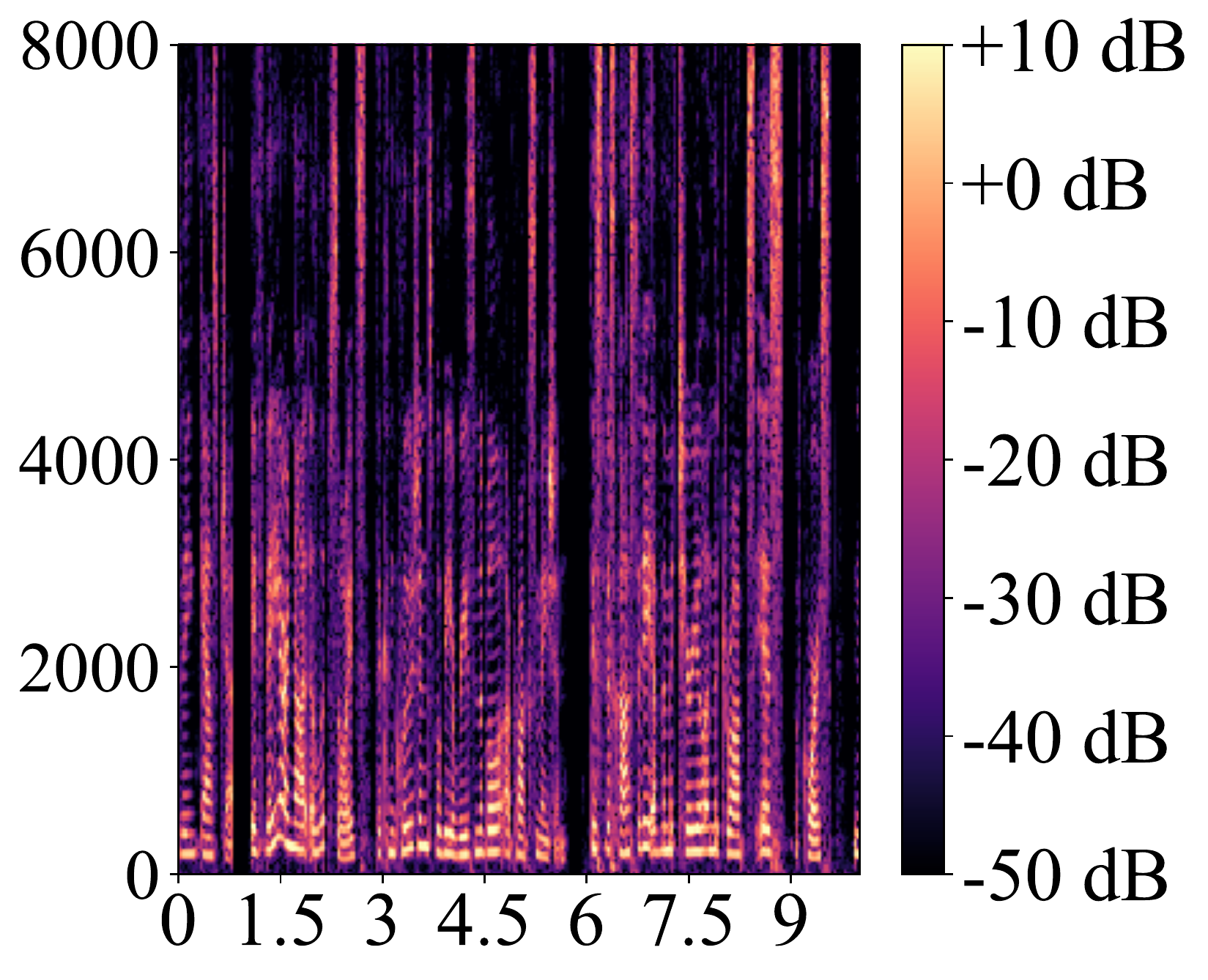}}
  \centerline{(b) Clean}
\end{minipage}
\begin{minipage}[b]{0.5\linewidth}
  \centering
  \centerline{\includegraphics[width=4cm, height=3.6cm]{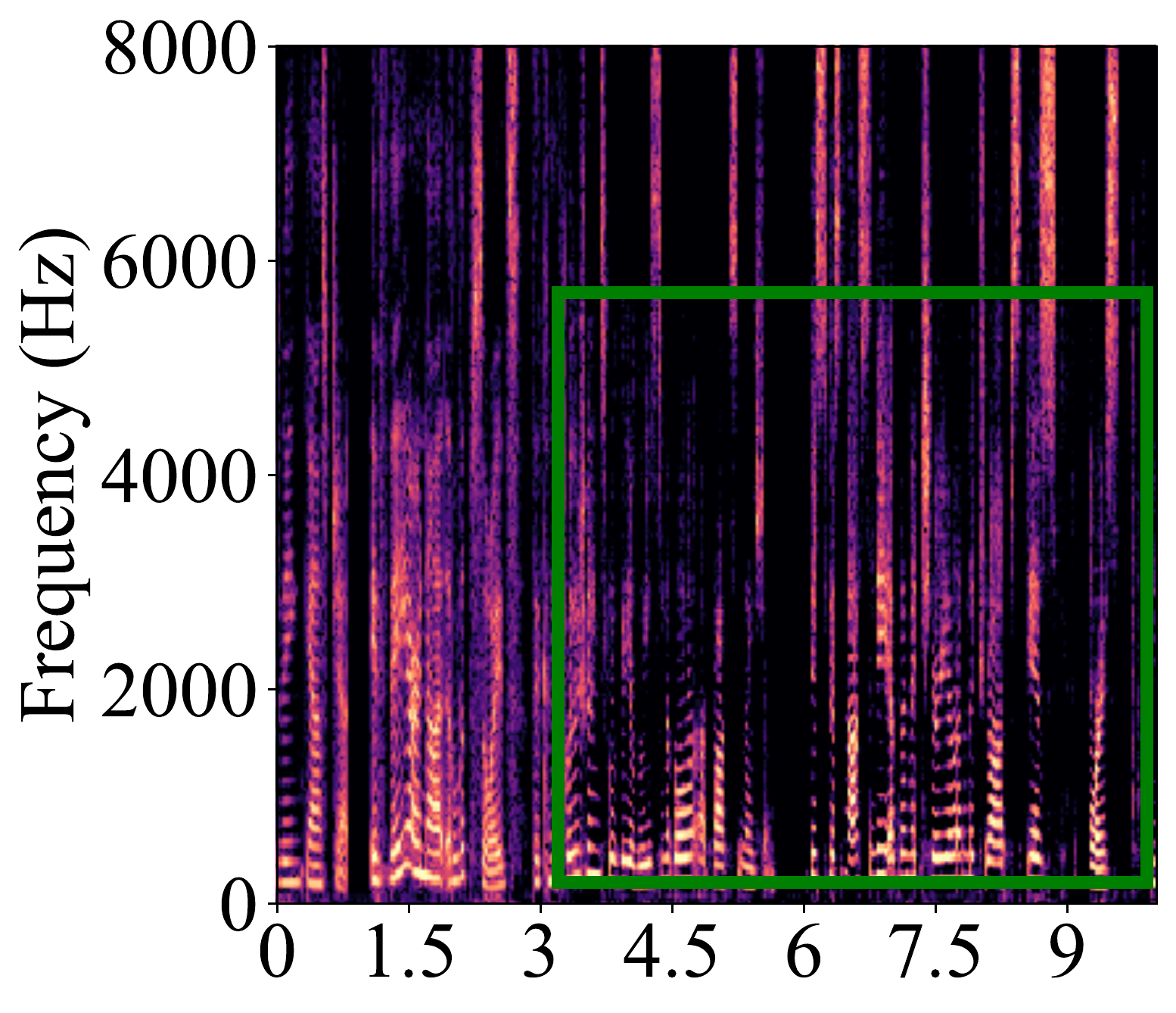}}
  \centerline{(c) WF}
\end{minipage}
\begin{minipage}[b]{.5\linewidth}
  \label{fig:error}
  \centering
  \centerline{\includegraphics[width=4cm, height=3.6cm]{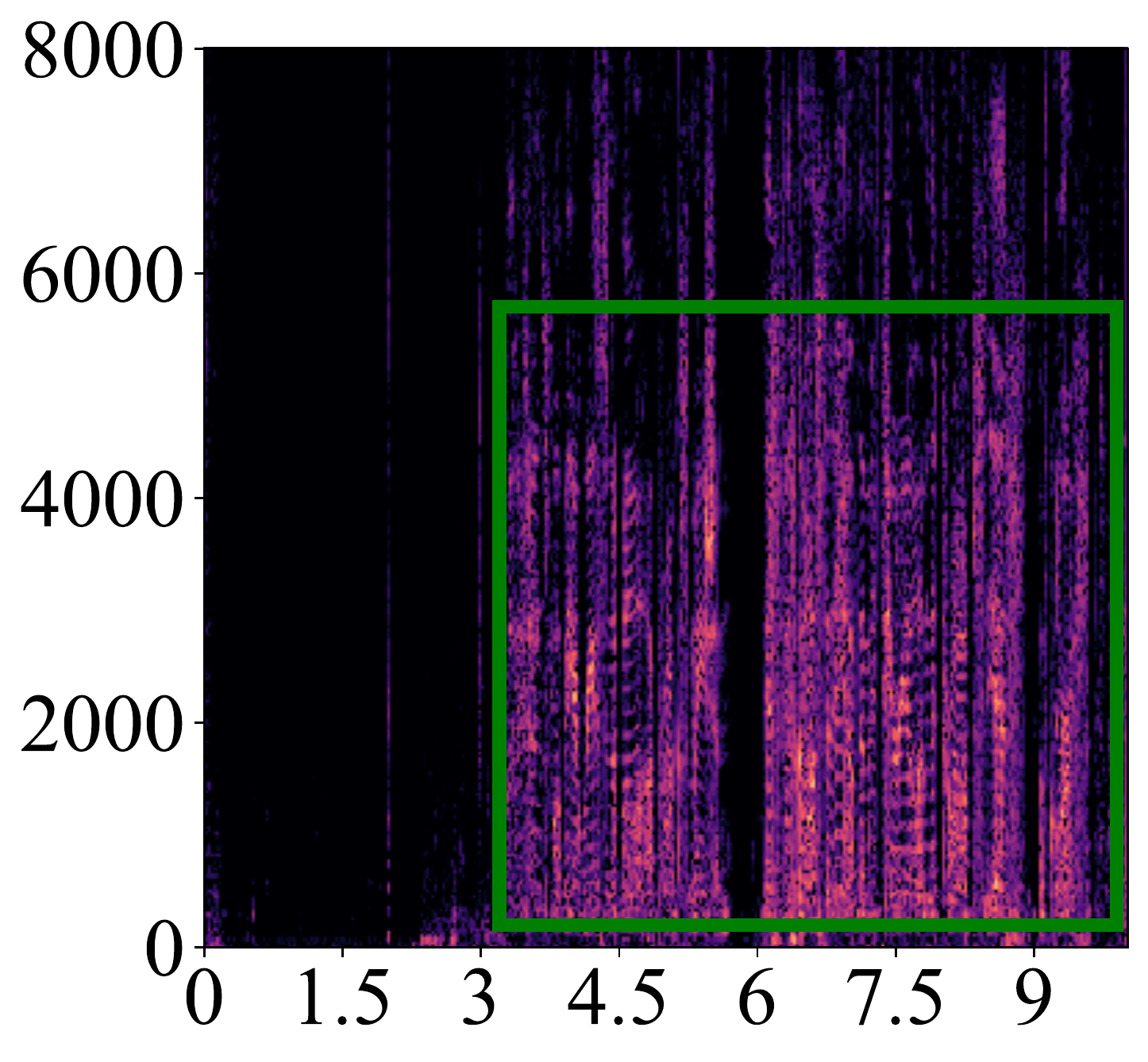}}
  \centerline{(d) Error}
\end{minipage}

\begin{minipage}[b]{0.5\linewidth}
  \centering
  \centerline{\includegraphics[width=4cm, height=3.7cm]{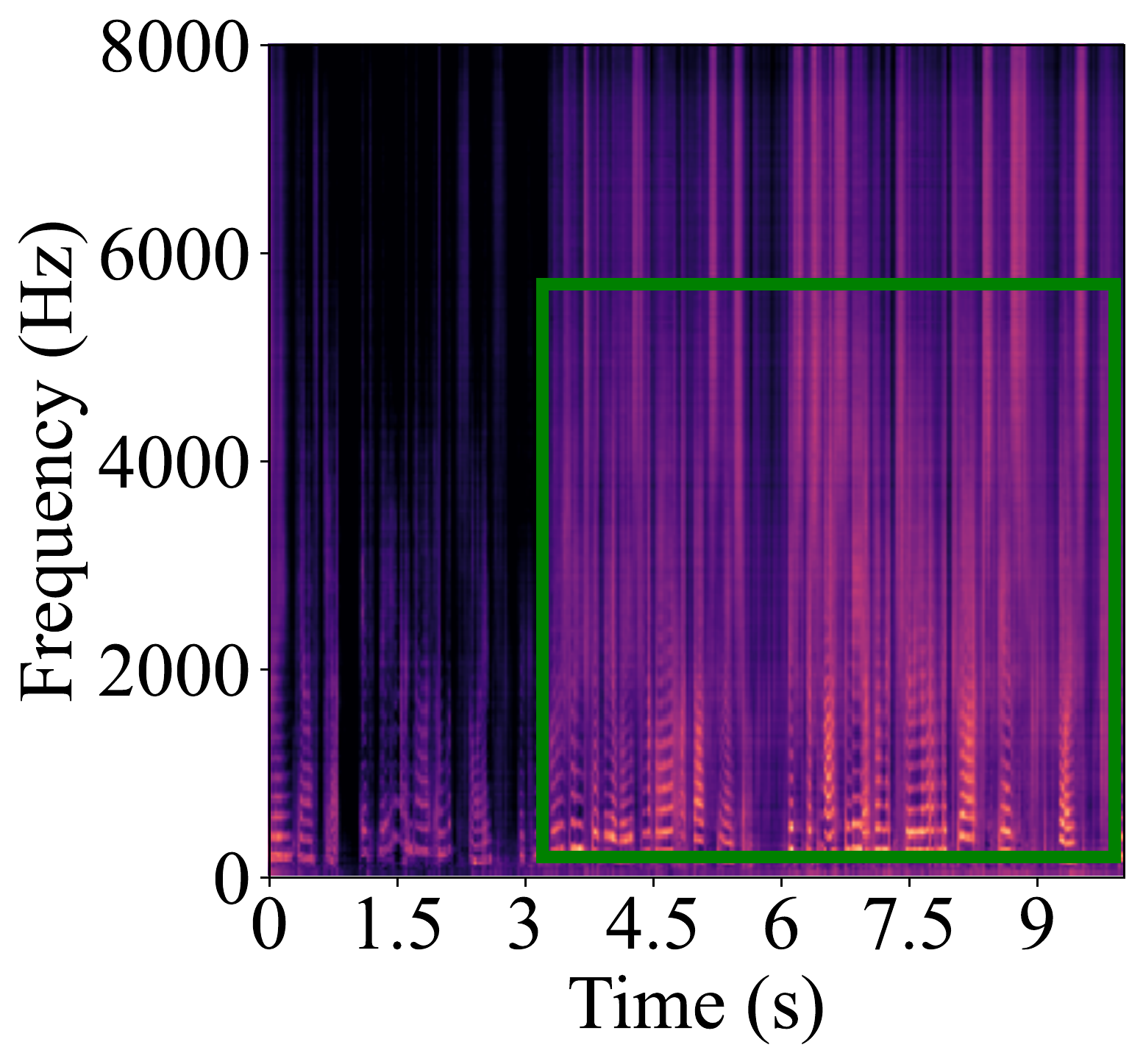}}
  \centerline{(e) Uncertainty}
\end{minipage}
\begin{minipage}[b]{0.5\linewidth}
  \centering
  \centerline{\includegraphics[width=4cm, height=3.7cm]{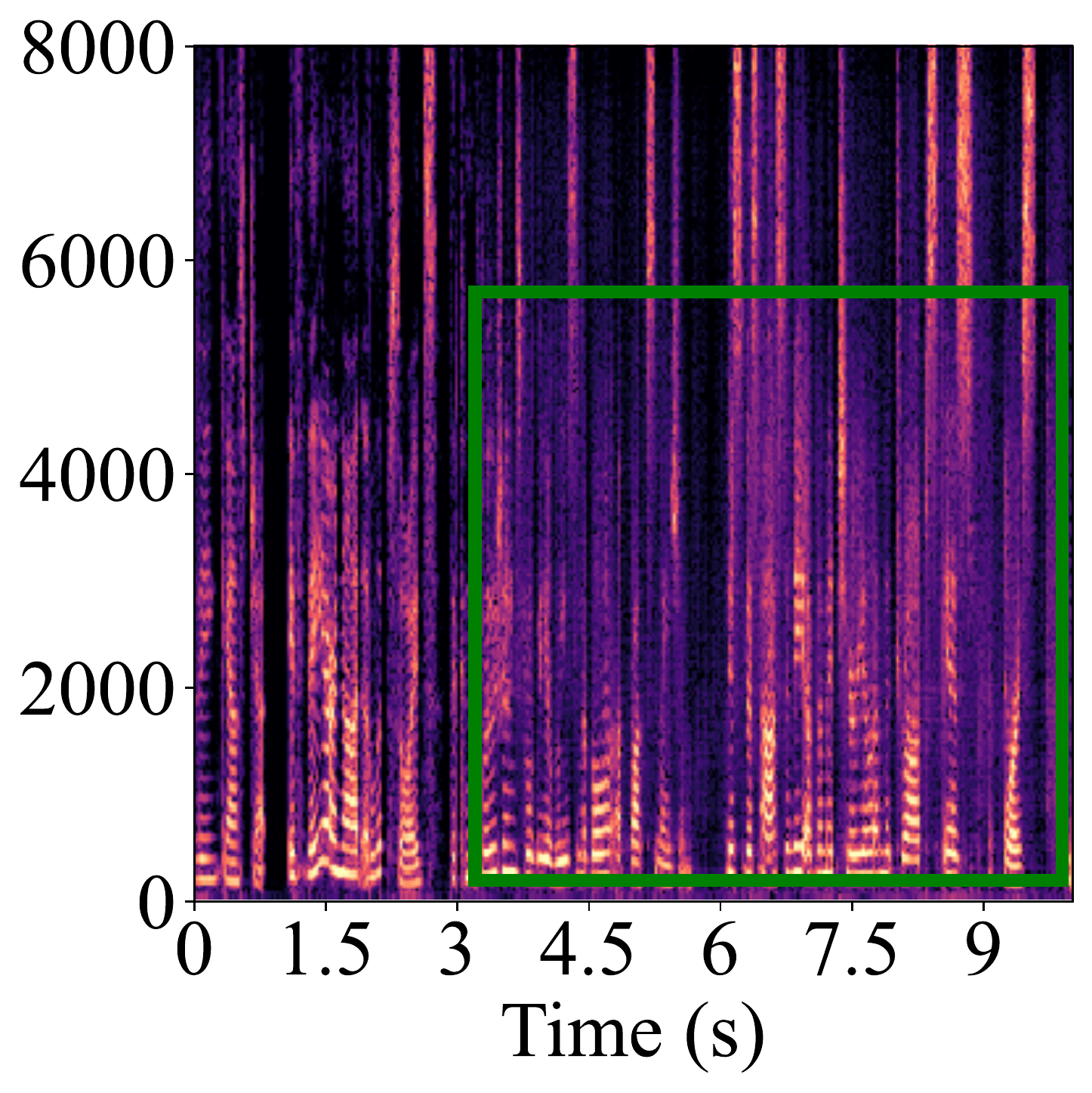}}
  \centerline{(f) A-MAP}
\end{minipage}
\caption{Example of estimation uncertainty captured by the proposed method on the DNS test dataset, shown in (e). The proposed method allows estimating clean speech by either using the estimated Wiener filter or applying the A-MAP estimator that incorporates both the estimated Wiener filter and the associated uncertainty, and the resulting estimates are shown in~(c) and~(f), denoted by WF and A-MAP, respectively. The estimation error of Wiener filtering in~(d) is computed between the estimated magnitudes~(c) and clean magnitudes~(b), indicating over- or under-estimation of speech magnitudes.}
\label{fig:illustraionofuncertainty}
\vspace{-0.3cm}
\end{figure}
\begin{figure*}[th]
\begin{minipage}[b]{0.33\linewidth}
  \centering
  \centerline{\includegraphics[width=6.3cm]{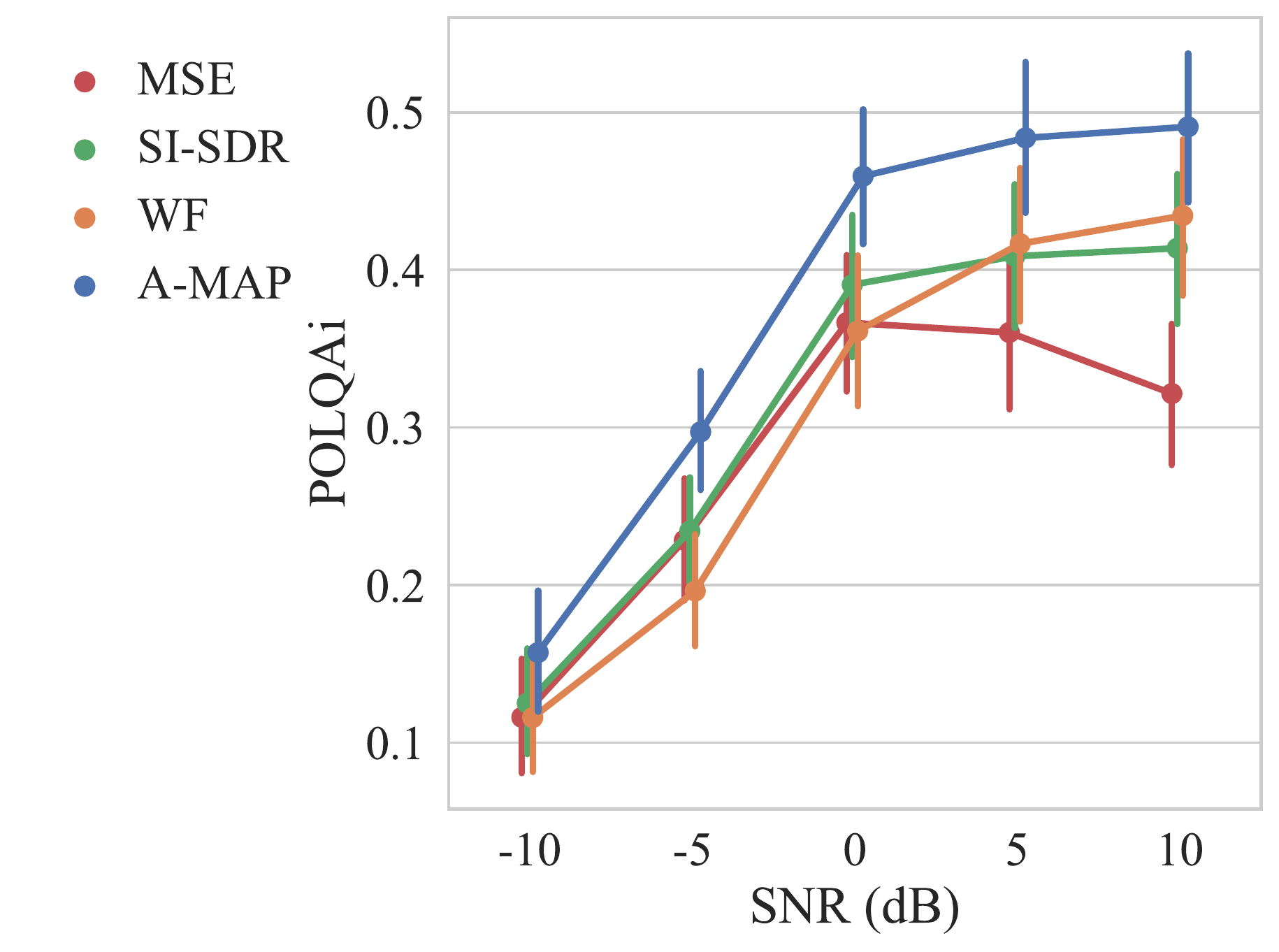}}
\end{minipage}
\begin{minipage}[b]{0.36\linewidth}
  \centering
  \centerline{\includegraphics[width=5cm]{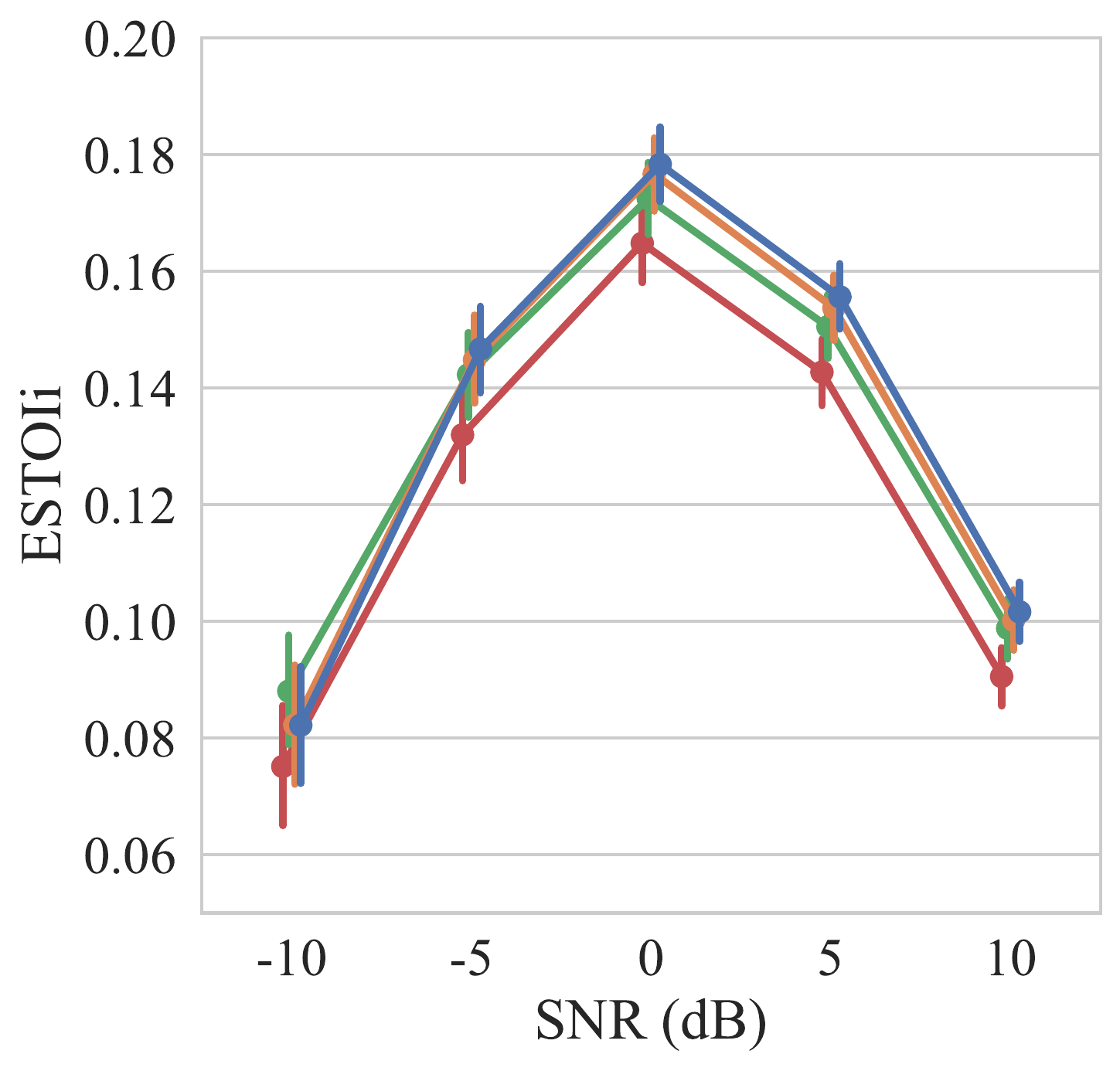}}
\end{minipage}
\begin{minipage}[b]{0.27\linewidth}
  \centering
  \centerline{\includegraphics[width=4.85cm]{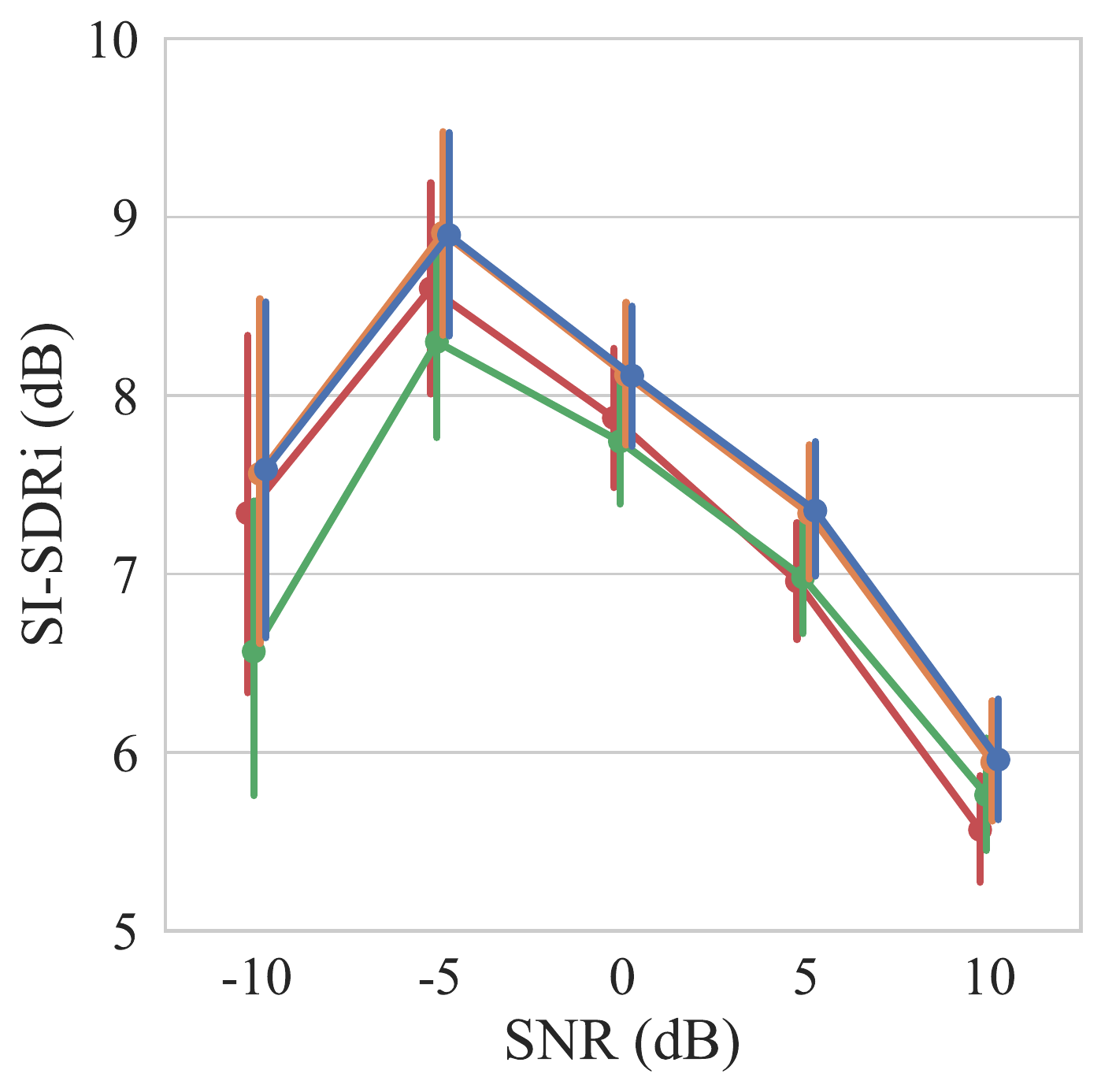}}
\end{minipage}
\caption{Performance improvement obtained on the synthetic dataset using clean speech from WSJ0 and noise signals from CHiME. POLQAi denotes POLQA improvement relative to noisy mixtures. The same definition applies to ESTOIi and SI-SDRi. The marker denotes the mean value over all utterances and the vertical bar indicates the 95\%-confidence interval.}
\label{fig:wsjnchime}
\vspace{-0.1cm}
\end{figure*}

 \vspace{-0.15cm}
\subsection{Baselines}
To evaluate the effectiveness of modeling uncertainty in neural network-based speech enhancement, we consider training the same neural network using standard cost functions, i.e., the \ac{MSE} defined as $\mathcal{L}_{\text{MSE}}$ in~\eqref{eqn:mse} and the SI-SDR defined as $\mathcal{L}_\text{SI-SDR}$ in~\eqref{eq:sisdr}. 
They are represented by MSE and SI-SDR in Table~\ref{dns_evaluation} and Fig.~\ref{fig:wsjnchime}. 
\vspace{-0.15cm}
\subsection{Hyperparameters}
All audio signals are sampled at 16~kHz and transformed into the time-frequency domain using the \ac{STFT} with a 32~ms Hann window and 50\% overlap.

For a fair comparison, we used the separator of Conv-TasNet~\cite{convtasnet2019} that has a~\ac{TCN} architecture. It has been shown to be effective in modeling temporal correlations. We used the causal version of the implementation and default hyperparameters provided by the authors\footnote{\url{https://github.com/naplab/Conv-TasNet}} without performing a hyperparameter search. Note that for our model performing uncertainty estimation, the output layer is split into two heads that predict both the Wiener filter and the uncertainty. We applied the sigmoid activation function to the estimated mask, while using the \emph{log-exp} technique to constrain the uncertainty output to be greater than $0$, i.e., the network outputs the logarithm of the variance, which is then recovered by the exponential term in the loss function. All neural networks were trained for 50 epochs with a batch size 16, the maximum norm of gradients was set to 5, and the parameters were optimized using the Adam optimizer~\cite{adamkinma} with a learning rate of 0.001. We halved the learning rate if the validation loss did not decrease for 3 consecutive epochs. To prevent overfitting, training was stopped if the validation loss failed to decrease within 10 consecutive epochs. The weighting factor $\beta$ is set to 0.01, chosen empirically.

\begin{table}[t!]
\centering
\resizebox{\columnwidth}{!}{
\begin{tabular}{|c||c||c|c|c|c|c|}
\hline
& POLQA & ESTOI & SI-SDR (dB)\\
\hline
Noisy & 2.30 $\pm$ 0.10 & 0.81 $\pm$ 0.02 & 9.07 $\pm$ 0.89\\
 \hline
 SI-SDR & 2.93 $\pm$ 0.11 & 0.88 $\pm$ 0.01 & 15.99 $\pm$ 0.75 \\
 MSE
 & 2.88 $\pm$ 0.10 & 0.88 $\pm$ 0.01 & 16.05 $\pm$ 0.71 \\
 \hline
Proposed WF
&  3.00 $\pm$ 0.11  & 0.88 $\pm$ 0.01  & 16.39 $\pm$ 0.73  \\
Proposed A-MAP
 &  \bftab 3.06 $\pm$ 0.10  & \bftab 0.89 $\pm$ 0.01  & \bftab 16.42 $\pm$ 0.73  \\
\hline
\end{tabular}
}
  \caption{Average performance over all utterances of the DNS non-reverberant synthetic test dataset in terms of POLQA, ESTOI, and SI-SDR. Values are given in mean $\pm$ confidence interval (95\% confidence).}
  \label{dns_evaluation}
  \vspace{-0.2cm}
\end{table}
\vspace{-0.15cm}
\section{Results and discussion}
\label{sec:results and discussion}
\subsection{Analysis of uncertainty estimation}
In Fig.~\ref{fig:illustraionofuncertainty}, we use an audio example from the \ac{DNS} test dataset to illustrate the uncertainty captured by the proposed method, and all plots are shown in decibel~(dB) scale. Applying the estimated Wiener filter to the noisy coefficients yields an estimate of the clean speech, denoted as WF shown in Fig.~\ref{fig:illustraionofuncertainty}~(c). To measure the prediction error, we can compute the absolute values of the difference between the estimated magnitudes, i.e., WF, and reference magnitudes given in Fig.~\ref{fig:illustraionofuncertainty}~(b), which indicates over- or under-estimation of speech magnitudes, shown in Fig.~\ref{fig:illustraionofuncertainty}~(d). It is observed that the model produces large errors when speech is heavily corrupted by noise, as can be seen by comparing the marking regions~(green boxes) of the noisy mixture shown in Fig.~\ref{fig:illustraionofuncertainty}~(a) and the prediction error of Fig.~\ref{fig:illustraionofuncertainty}~(d). By comparing error in Fig.~\ref{fig:illustraionofuncertainty}~(d) and  uncertainty in Fig.~\ref{fig:illustraionofuncertainty}~(e), the estimator generally associates large uncertainty  with large prediction errors, while giving low uncertainty to accurate estimates, e.g., the first 3~seconds. This shows that the model produces uncertainty measurements that are closely related to estimation errors. In our proposed method with uncertainty estimation, we can use not only the estimated Wiener filter, but also the estimated \ac{A-MAP} mask that incorporates both the estimated uncertainty and Wiener filter, as given in \eqref{eqn:approximated_map}. This estimate is denoted as \ac{A-MAP} in Fig.~\ref{fig:illustraionofuncertainty}~(f). We observe that the \ac{A-MAP} estimate causes less speech distortion compared with the WF estimate, as can be seen, e.g., from the marking regions of WF and A-MAP.
\vspace{-0.15cm}
\subsection{Performance Evaluation}
In Table~\ref{dns_evaluation}, we present average evaluation results of our method on the \ac{DNS} synthetic test set in terms of \ac{SI-SDR} measured in dB, \ac{ESTOI}~\cite{estoi}, and \ac{POLQA}\footnote{We  would like to thank  J. Berger and Rohde\&Schwarz SwissQual AG for their support with POLQA.}~\cite{polqa}. We observe that modeling uncertainty yields improvement over the baselines, where the proposed WF outperforms the baselines in terms of \ac{POLQA} and \ac{SI-SDR}, and a larger improvement can be observed between the baselines and the proposed A-MAP.  This shows that it is advantageous to model uncertainty within the model instead of directly estimating optimal points.

In Fig.~\ref{fig:wsjnchime}, we present speech enhancement results in terms of mean improvement of \ac{POLQA}, \ac{ESTOI}, and \ac{SI-SDR}. For this evaluation we used another unseen test dataset based on speech from WSJ0 and noise from CHiME. It shows that our proposed approach performs better in terms of speech quality given by higher \ac{POLQA} values without deteriorating \ac{ESTOI} (with an exception at SNR of $-10$~dB) and \ac{SI-SDR}, which again demonstrates the benefit of modeling uncertainty. We also observe that larger improvement over the baselines is achieved at high \acp{SNR}, which may be explained by the fact that, at high \acp{SNR}, speech quality (and thus \ac{POLQA}) is mainly affected by speech distortions, while at low \acp{SNR} the main factor is residual noise.
\vspace{-0.05cm}
\section{Conclusion}
\label{sec:conclusion}
\vspace{-0.05cm}
Based on the common complex Gaussian model of speech and noise signals, we proposed to augment the existing neural network architecture with an additional uncertainty estimation task. Specifically, we proposed simultaneous estimation of the Wiener filter and the associated uncertainty to capture the full speech posterior distribution. Furthermore, we proposed using the estimated Wiener filter and uncertainty to produce an A-MAP estimate of the clean spectral magnitude. Eventually, we combined uncertainty estimation and speech enhancement by the proposed hybrid loss function. We showed that the approach can capture uncertainty and lead to improved speech enhancement performance across different speech and noise datasets. For future work, it would be interesting to integrate the uncertainty estimation into multi-modal learning systems, which may rely more on other modalities when audio modality raises high uncertainty.

\AtNextBibliography{\small}
\section{REFERENCES}
\label{sec:refs}
\atColsBreak{\vskip5pt}
\printbibliography[heading=none]
\end{document}